# Fault Tolerant Processing Unit Using Gamma Distribution Sliding Window For Autonomous Landing Guidance System

Hossam O. Ahmed,
College of Engineering and Technology,
*American University of the Middle East, Kuwait.*

*Abstract*—To keep up with today's dense metropolitan areas and their accompanying traffic problems, a growing number of towns are looking for more advanced and swift urban taxi drones. The safety parameters that must be taken into consideration may be the most important element in the widespread use of such technology. Most recent aviation mishaps have happened during the landing phase, making this a particularly important safety consideration for Vertical and/or Short Take-Off and Landing (V/STOL) drones.

In this study, we focused on improving the fault tolerance of the processor architectures used by the predecessors of Autonomous Landing Guidance Assistance Systems (ALGAS), which in turn improves their decision-making capabilities. Furthermore, this is achieved by proposing a fault-tolerant processing architecture that depends on the Gamma Distribution Sliding Window Unit (GDSWU). This proposed GDSWU has been designed completely using VHDL, and the targeted FPFA was the Intel Cyclone V 5CGXFC9D6F27C7 chip. The GDSWU could operate at a maximum frequency of 369.96 MHz, as calculated by the synthesis results of the INTEL Quartus Prime program. The suggested GDSWU core only requires 20.36 mW for dynamic core and I/O power consumption.

## I. INTRODUCTION

Existing transportation infrastructure will be severely overburdened by the world's ever-increasing population, and drastic measures will be required. There have been several efforts and ideas to distribute air corridors at various levels. Depending on promising avionics technology such as Urban Air Mobility (UAM) is considered a logical option, especially for drones that use vertical takeoff and landing (V/STOL) [1-3]. Due to the massive UAM taxi drones that are predicted to span each metropolis in the near future, the integrated V/STOL mechanisms must be very robust against unforeseen and foreseeable problems [4-5].

Also, it's important to keep in mind that the last three stages of every aircraft flight—the first approach, final approach, and landing—account for around 53.85% of all modern flight mishaps [6, 7]. In this paper, we propose to support the previous fault-tolerant processing architectures, for Autonomous Landing Guidance Assistance Systems (ALGAS), with the Gamma Distribution Sliding Window Unit (GDSWU). This will improve the overall decision-making capabilities as will be explained in the following Sections.

## II. PRELIMINARY

Successful global implementation of UAM requires a deep comprehension of and proficiency in implementing safety criteria. Current safety requirements for air transportation systems like Single European Sky ATM Research (SESAR) and Next Generation (NextGen) should be merged with new, flexible safety standards. Increasingly, the fuzzy logic system (FLS) has been presented as an enhanced control method to solve several UAM concerns, such as autopilot dynamics and visual human tracking in drone systems. The FLS can also manage electrical system signal ambiguity and noise [8-9].

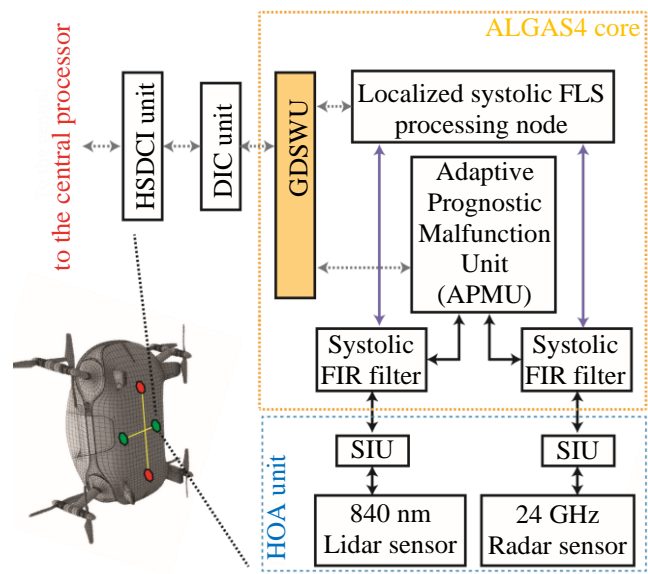

Fig. 1. The detailed structure of one spatial corner of the proposed ALGAS4 system with the GDSWU.

## III. THE PROPOSED GDSWU PROCESSING CORE

This paper is a continuation of the proposed ALGAS Processing system for UAM drones [8-9]. It is enhancing the safety feature of V/STOL taxi drones during the landing mission. The proposed ALGAS system consists of four ALGAS processing cores. Each ALGAS processing core is illustrated in Fig.1. The main task of each ALGAS processing node is to continuously measure the sensory data from the short-range radar and the lidar sensors via the Sensor Interface Unit (SIU). Then, these data are filtered from noises using the two FIR filters. The outputs of the FIR filters are further processed using the Adaptive Prognostic Malfunction Unit (APMU) and the FLS unit. The APMU is responsible for indicating the agility degree between the two sensory signals by measuring the discrepancy rate between their readings. The FLS is responsible for measuring the altitude between the drone and the landing surface. All these blocks have been explicitly described in the previous publications [8-9]. The output signal from the FLS provides crucial information on which the stability of the entire system is depending on during a landing operation. Thus, a single error from the FLS could cause dramatic negative consequences that the drone's actuators could not reverse it. Hence, this paper proposed the GDSWU to enhance the fault tolerance capability of the proposed ALGAS system and support durable information to the decision-making provider. Moreover, the final decision from each ALGAS core will not only depend on the current response of the FLS unit but also on the accumulated previous 16 consecutive outputs as well. Each of these previous output samples is multiplied by a certain weight according to the Gamma Distribution function as shown in (1); where $\Gamma(a)$ is the Gamma function under the condition that all positive integers $\Gamma(a)=(a-1)!$.

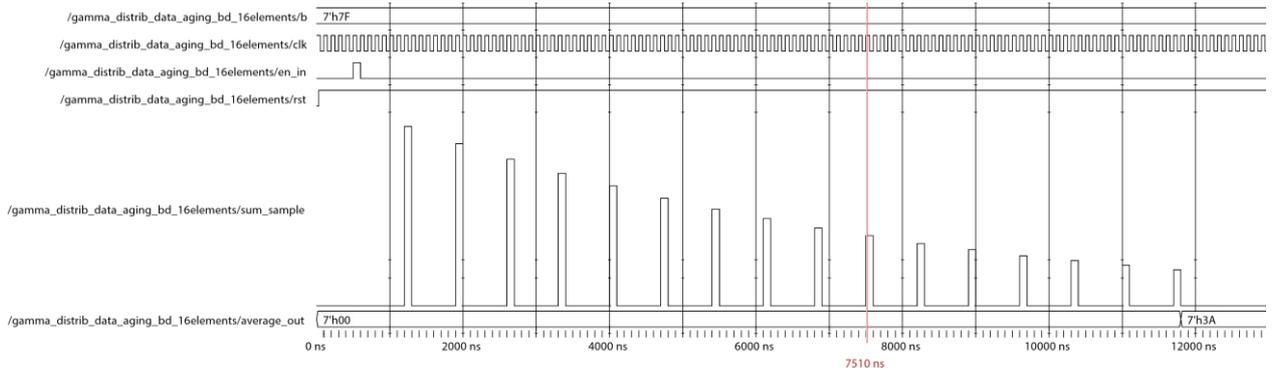

Fig. 2. The waveform of the proposed 16-tap gamma distribution sliding window.

$$f(x; a, b) = \frac{1}{b^a \Gamma(a)} x^{a-1} e^{\frac{-x}{b}} \quad (1)$$

TABLE I. THE SUMMARY OF THE COMPUTATIONAL PERFORMANCE OF THE PROPOSED 16-TAP GAMMA DISTRIBUTION SLIDING WINDOW.

|  | The proposed 16-tap gamma distribution sliding window in this paper |
|---|---|
| *Targeted FPGA device name* | INTEL 5CGXFC9D6F27C7 |
| *Embedded FPGA's DSP resources usage* | None |
| *Total dynamic core and I/O thermal power dissipation* | 20.36 mW |
| *External memory usage or BRAM* | None |
| *Max. operating frequency* | 369.96 MHz |
| *Number of logic elements* | 464 ALMs |
| *Total number of registers* | 922 registers |
| *Number of parallel operations per clock cycle* | 22 operations/clock cycle |
| *Reconfigurable gamma distribution parameters* | Fixed, a=1 and b=10 |

## IV. THE RESULTS

The proposed GDSWU has been designed using VHDL at the Gate and RTL levels without the usage of any IPs to optimize the computing performance. We selected a=1 and b=10. As illustrated in Fig.2, the waveform using Questa Simulation shows the step response of the GDSWU unit. The average_out signal shows the final accumulated average value per the sliding window. As depicted in Table I, INTEL Quartus Prime's synthesis shows that the coarse-grained architecture of the GDSWU can operate at a maximum frequency of 369.96 MHz. The dynamic core and I/O power consumption of the proposed GDSWU system was just 20.36 mW. From Fig. 2, we could notice that if the initial seed is 7h'7F, then the final accumulated average after the 16 taps will be 7h'3A. The Gamma distribution could be changed based on the experts and the targeted application. The ultra-low power consumption of the GDSWU is due to many factors of optimization and the usage of variable-width unsigned number representation for processing the data over the proposed systolic processing architecture. In addition to the enhanced fault-tolerance capability, the proposed GDSWU will also improve the overall computation performance of the ALGAS core [9] by adding 22 operations per clock cycle.

## V. CONCLUSION

The proposed GDSWU provides more agile output signals. This feature will improve the overall fault tolerance of the futuristic Vertical and/or Short Take-Off and Landing (V/STOL) drones. The next version of the proposed GDSWU will have reconfigurable parameters according to the preferable landing scenario.